\documentclass[lettersize,journal]{IEEEtran}
\usepackage{amsmath,amsfonts}
\usepackage{algorithmic}
\usepackage{array}
\usepackage[caption=false,font=normalsize,labelfont=sf,textfont=sf]{subfig}
\usepackage{textcomp}
\usepackage{stfloats}
\usepackage{url}
\usepackage{verbatim}
\usepackage{graphicx}
\usepackage{setspace}
\usepackage{multirow}
\usepackage[table]{xcolor}
\usepackage{tabularx} 
\usepackage{cite}
\usepackage{makecell}
\usepackage[center]{caption}
\usepackage{pict2e,picture}
\newsavebox\CBox
\newlength\CLength
\def\Circled#1{\sbox\CBox{#1}%
  \ifdim\wd\CBox>\ht\CBox \CLength=\wd\CBox\else\CLength=\ht\CBox\fi
    \makebox[1.5\CLength]{\makebox(0,1.5\CLength){\put(0,0){\circle{1.5\CLength}}}%
    \makebox(0,1.5\CLength){\put(-.5\wd\CBox,0){#1}}}}

\def\BibTeX{{\rm B\kern-.05em{\sc i\kern-.025em b}\kern-.08em
    T\kern-.1667em\lower.7ex\hbox{E}\kern-.125emX}}
\usepackage{balance}
\begin{document}
\title{Forward and Reverse Converters for the Moduli-Set $\{2^{2q+1},2^q+2^{q-1}\pm1\}$ }
\author{Ghassem Jaberipur, Bardia Nadimi, R. Kazemi, and Jeong-A Lee, Member, IEEE

\thanks{G. Jaberipur is with the Brain Pool Program for Chosun University, Gwangju, South Korea (e-mail: Jaberipur@chosun.ac.kr). B. Nadimi is a Ph.D. student at the University of South Florida, Tampa, Florida, USA (e-mail: bnadimi@usf.edu). R. Kazemi is a graduate of Computer Engineering from the Department of Computer Science and Engineering, Shahid Beheshti University, Tehran, Iran (e-mail: R.Kazemi@mail.sbu.ac.ir). Jeong-A Lee is with the Department of Computer Engineering, Chosun University, Republic of Korea (e-mail: Jalee@chosun.ac.kr).}
}

\markboth{Journal of \LaTeX\ Class Files,~Vol.~18, No.~9, September~2020}%
{How to Use the IEEEtran \LaTeX \ Templates}

\maketitle

\begin{abstract}

Modulo-$(2^q+2^{q-1}\pm1)$ adders have recently been implemented using the regular parallel prefix (RPP) architecture, matching the speed of the widely used modulo-$(2^q\pm1)$ RPP adders. 
Consequently, we introduce a new moduli set $\tau^+=\{2^{2q+1},2^q+2^{q-1}\pm1\}$, with over $(2^{q+2})\times$ dynamic range and adder speeds comparable to the conventional $\tau=\{2^q,2^q\pm1\}$ set. 
However, to fully leverage $\tau^+$ in residue number system applications, a complete set of circuitries is necessary. 
This work focuses on the design and implementation of the forward and reverse converters for $\tau^+$. 
These converters consist of four and seven levels of carry-save addition units, culminating in a final modulo-$(2^q+2^{q-1}\pm1)$ and modulo-$(2^{2q+1}+2^{2q-2}-1)$ adder, respectively. 
Through analytical evaluations and circuit simulations, we demonstrate that the overall performance of a sequence of operations—including residue generation, $k$ additions, and reverse conversion—using $\tau^+$ surpasses that of $\tau$ when $k$ exceeds a certain practical threshold.
\end{abstract}


\section{Introduction}
\IEEEPARstart{T}{HE} decades-long classical moduli-set $\tau=\{2^q,2^q\pm1\}$ has been frequently used in numerous applications of residue number systems (RNS), including the popular deep neural network hardware accelerators \cite{Application1, Application2, Application3}, FIR filters \cite{Application4, Application5}, and image processing \cite{Application6, Application7, Application8}. 
However, on demand for additional dynamic range (DR), one has two options; namely 1) Increase the channel widths $q$, to the extent that additional delay is tolerable. 2) Introduce additional $\tau$- balanced moduli (i.e., augmenting $\tau$ with some moduli, for which the same speed arithmetic operations are possible). 
In other words, as a necessary but not essentially sufficient condition, the residues of any added moduli should be representable with roughly the same $q$ bits, as of the original moduli $\tau$, as higher values of $q$ will lead to increased delays and make the moduli set imbalance.  
For example, the two conjugate moduli $2^q+2^{q-1}\pm1$, have been recently studied in \cite{OurRNSPaper} and \cite{DaryushPaper}, where the corresponding modular adders, and that of $2^{2q+1}$, with parallel prefix architectures, show to be equally fast as the similar modulo-$(2^q\pm1)$ adders in regular parallel prefix (RPP, as named in \cite{ref11}) realizations \cite{ref12, ref13, ref14, ref15}. 
Therefore, $\tau^+=\{2^{2q+1},2^q+2^{q-1}\pm1\}$ can be used as efficiently as the popular $\tau$, with the benefit of $(2^{q+2})$X DR. 

However, for a truly effective use in the prospective applications, one needs to design and implement balanced multipliers with those of $\tau$, and provide for reasonably efficient forward and reverse converters for $\tau^+$. 

In this work, we offer the required converters and leave the multipliers for another ongoing project. 
The remainder of this work contains the following sections. 
We provide a brief introduction to RNS essentials, including the Chinese remainder theorem (CRT) in the next section. 
Details of forward and reverse converters for $\tau^+$ can be found in Section III, and analytical gate-level evaluation and the results of circuit simulation and synthesis are in Section IV. 
Finally, see the concluding remarks in Section V.

\section{RNS Essentials}
Binary numbers $X$, $Y$, and $Z$ are represented in a $k$-moduli RNS $\{m_1,\cdots, m_k\}$, as $X=(x_1,\cdots,x_k)$, $Y=(y_1,\cdots,y_k)$, and $Z=(z_1,\cdots,z_k)$, where $x_i=|X|_{m_i}$, $y_i=|Y|_{m_i}$, and $z_i=|Z|_{m_i}$, denote the remainder of integer divisions $X/m_i$, $Y/m_i$, and $Z/m_i$, for $1\leq i\leq k$, respectively.  
The most viable RNS operations are $+$ and $\times$, where the RNS equivalent of binary operations $Z=X+Y$ and $Z=X\times Y$, are obtained as $z_i=|x_i+y_i|_{m_i}$, and $z_i=|x_i \times y_i|_{m_i}$, respectively.

Extracting the residues from the binary operand is called residue generation (aka forward conversion). 
This simple operation is commonly done in parallel for all the $k$ moduli. 
However, the reverse conversion, as a function of all the $k$ residues, is often hard to implement. 
This is through one or a mixture of alternative versions of the Chinese remainder theorem (CRT); namely the plain CRT, New CRT \cite{ref16}, and mixed radix conversion \cite{ref17}, where the latter is a sequential operation manipulating the residues one by one. 
Equation (\ref{eqn1}) provides for the New CRT, where  $\mu _i=p_{2-i} \hat{p_{1-i}}$, $p_{2-i}=\prod_{j=2}^im_j$, $\hat{p_{1-i}}$ is the multiplicative inverse of $p_{1-i}=\prod_{j=1}^im_j$  (i.e., $|p_{1-i}\hat{p_{1-i}}|_{p_{i+1-k}}=1$), $p_{i+1-k}=\prod_{j=i+1}^km_j$  $(1\leq i<k)$, $x_i=|X|_{m_i}$, and $M_1=M/m_1$. 

\begin{equation}
\label{eqn1}
    X=x_1+m_1 |\sum_{i=1}^{k-1} \mu _i (x_{i+1}-x_i ) |_{M_1}
\end{equation}

For example, (\ref{eqn2}) describes the New CRT formula for $\tau$, where $\mu _1=2^q$ and $\mu _2=2^{q-1}+1$.

\begin{equation}
\label{eqn2}
    X=x_1+2^q |\mu _1 (x_2-x_1 )+\mu _2 (x_3-x_2 )|_{2^{2q}-1}
\end{equation}
\section{Forward and reverse converters for $\tau^+$}
Forward conversion for the commonly utilized moduli $2^q\pm 1$, is straightforward, employing simple circuit designs as detailed in \cite{ref11, ref14}. 
However, designing residue generators for the more general case of $2^q \pm \delta$ where $3\leq \delta \leq 2^{q-1}+1$ (including $2^q+2^{q-1}\pm1$) has not been extensively explored in the literature. 
This is despite the fact that efficient modular adders have been developed for $\delta=2^k\pm1$, where $k\leq q-1$ (including $2^q+2^{q-1}\pm1$) \cite{ref18, ref19}.
Therefore, there is a clear motivation to study residue generators for $\tau^+$.

\vspace*{-10pt}
\subsection{Residue generators (forward converters) for $\tau ^+$}
The DR of $\tau ^+=\{m_1=2^{2q+1},m_2=2^q+2^{q-1}-1,m_3=2^q+2^{q-1}+1\}$ covers $(4q+3)$-bit integers in $[0,2^{4q+2}+2^{4q-1}-2^{2q+1})$ that can be represented as in (\ref{eqn3}), where $X_3=0x_{4q+2}\cdots x_{3q+3}$, $X_2=x_{3q+2}\cdots x_{2q+2}$, $X_1=x_{2q+1}\cdots x_{q+1}$, and $X_0=x_q\cdots x_0$. 

\begin{equation}
\label{eqn3}
    X=2^{3q+3} X_3+2^{2q+2} X_2+2^{q+1} X_1+X_0
\end{equation}

\begin{figure}
    \centering
    \captionsetup{justification=centering}
    \includegraphics[width=.8\columnwidth]{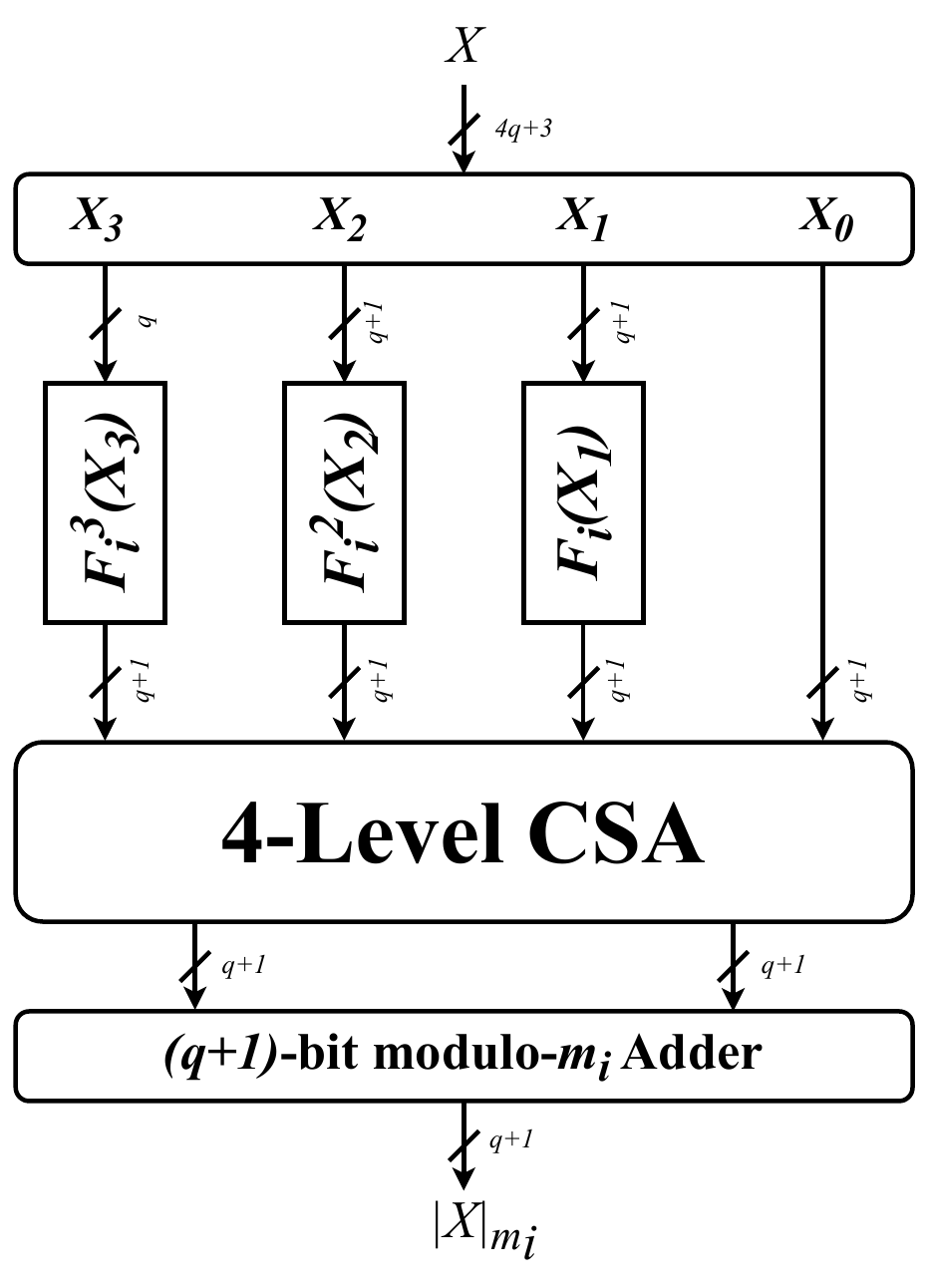}
    \caption{Modulo-$m_i$ forward converter $(i\in\{2,3\})$}
    \label{fig:forwardConverterFigure}
\end{figure}

Let $F_2 (Z)=|2^{q+1} Z|_{m_2}$ and $F_3 (Z)=|2^{q+1} Z|_{m_3}$ be implemented via $2^{q+1} (q+1)$-bit look-up tables (LUT), where $Z$ is a $(q+1)$-bit residue. Also, let $F_i^2 (Z)=F_i (F_i (Z))$, and $F_i^3 (Z)=F_i (F_i (F_i (Z)))$. 
Therefore, (\ref{eqn4}), and (\ref{eqn5}) are easily derived. 
Fig.~\ref{fig:forwardConverterFigure} depicts the required circuitry for $|X|_{m_2}$ and $|X|_{m_3}$, where the critical delay path travels through one LUT, four levels of CSAs, and one modulo-$m_i$ adder.

\vspace*{-10pt}
\begin{equation}
\label{eqn4}
    |X|_{m_1}=|2^{q+1}X_1+X_0|_{2^{2q+1}}=x_{2q}\cdots x_{q+1} x_q\cdots x_0
\end{equation}
\vspace*{-10pt}
\begin{equation}
\label{eqn5}
    |X|_{m_i}=|F_i^3 (X_3)+F_i^2 (X_2)+F_i (X_1)+X_0 |_{m_i}
\end{equation}

\vspace*{-10pt}
\subsection{Multiple-residue to binary (Reverse) converter for $\tau^+$}
We use the $\{m_1,\{m_2,m_3\}\}$ grouping of moduli, and apply the New CRT formula (\ref{eqn1}) on $\{m_2,m_3\}$, and on $\{m_1,m_2m_3\}$. However, after some tedious elaborations, we combine the two formulas leading to a more efficient implementation. Application of the New CRT on the moduli pair $m_2=2^q+2^{q-1}-1$, and $m_3=m_2+2$, leads to (\ref{eqn7}), as follows, where the auxiliary definitions {\large \textcircled{\normalsize a}} to {\large \textcircled{\normalsize h}} can be found in the Appendix A. 

\textbf{New CRT on $\{m_2,m_3\}$:}\\
$X_{23}=x_3+m_3 |\mu _2 (x_2-x_3)|_{m_2}=$\\$x_3+|\mu _2 m_3 (x_2-x_3)|_{m_2 m_3}$, per {\large \textcircled{\normalsize a}}. \\
Let $X'_{23}=|\mu _2 m_3 (x_2-x_3)|_{m_2 m_3}$, per {\large \textcircled{\normalsize d}} - {\large \textcircled{\normalsize e}} $\Rightarrow$ \\ 
$X'_{23}=|16\mu _2 m_3 (x'_2-x'_3)+\mu _2 m_3 (x''_2-x''_3)|_{m_2 m_3}=$ \\ $|8(m_2+1)m_3 (x'_2-x'_3)+\mu _2 m_3 (x''_2-x''_3)|_{m_2 m_3}$, per {\large \textcircled{\normalsize b}} - {\large \textcircled{\normalsize e}} $\Rightarrow$ 
\begin{equation}
\label{eqn7}
    X'_{23}=|8m_3 (x'_2-x'_3)+\mu _2 m_3 (x''_2-x''_3)|_{m_2 m_3}
\end{equation}

For the 2$^{nd}$ application of New CRT the following are used:

\textbf{New CRT on $\{m_1,m_2 m_3\}$:} 
\begin{spacing}{1.2}
\noindent $X=x_1+m_1 X', X'=|\mu _1 (X_{23}-x_1)|_{m_2 m_3}=$ \\ $|\mu _1 (x_3+X'_{23}-x_1)|_{m_2 m_3}=|X''_{23}+\mu _1 (x_3-x_1)|_{m_2 m_3}$, where $X''_{23}=|\mu _1 X'_{23}|_{m_2 m_3}$. \\
\noindent $X''_{23}=|(3\times 2^{q-4} (m_2+1)+1)X'_{23}|_{m_2 m_3}$, per {\large \textcircled{\normalsize h}} $\Rightarrow$ \\ $X''_{23}=|(3\times 2^{q-4}+1)X'_{23}|_{m_2 m_3}=$ \\ $|((3\times 2^{q-4}+1)(8m_3(x'_2-x'_3)+\mu _2 m_3(x''_2-x''_3)))|_{m_2 m_3}$, per (9) $\Rightarrow$ $X''_{23}=|((3\times 2^{q-1}+8) m_3 (x'_2-x'_3)+\mu _2 (3\times 2^{q-4}+1) m_3 (x''_2-x''_3))|_{m_2 m_3}= |((m_2+9) m_3 (x'_2-x_3)+(2\mu _2 (3\times 2^{q-5} )+\mu _2)m_3 (x''_2-x''_3))|_{m_2 m_3}$, per {\large \textcircled{\normalsize g}} $\Rightarrow$ 
\end{spacing}
\vspace*{-8pt}
\begin{equation}
\label{eqn8}
    X''_{23}=|9m_3 (x'_2-x'_3)+(3\times 2^{q-5}+\mu _2) m_3 (x''_2-x''_3)|_{m_2 m_3}
\end{equation}
\noindent $X'=|\mu _1 (x_3-x_1)+X''_{23}|_{m_2 m_3}=|(\mu _1 (16x'_3+x''_3)-\mu _1 x_1+9m_3 (x'_2-x'_3)+(3\times 2^{q-4}+1) m_3 (x''_2-x''_3))|_{m_2 m_3}$ \\
The latter leads to (\ref{eqn9}) as found below, via replacing the negative terms with their positive complementary equivalents. 
The corresponding tedious elaborations can be found in the Appendix B.\\
\vspace*{-8pt}
\begin{spacing}{1.2}
\noindent $x_1=|X|_{m_1}=8x_1^{'}+x_1^{''}$, $x_1^{'}=x_{1_2q}\cdots x_{1_3}$, $x_1^{''}=x_{1_2} x_{1_1} x_{1_0}$, $-x_1^{'}=\overline{x_1^{'}}+1-2^{2q-2}$, $\overline{x_1^{'}}=\overline{x_{1_2q}}\cdots \overline{x_{1_3}}$, $-x_1^{''}= \overline{x_1^{''}}-7$,$\overline{x_1^{''}}=\overline{x_{1_2}}\overline{x_{1_1}}\overline{x_{1_0}}$, $-x_3^{'}=\overline{x_3^{'}}+1-2^{q-2}$, $\overline{x_3^{'}}=\overline{x_{3_q} }\cdots \overline{x_{3_3}}$, $-x_3^{''}=\overline{x_3^{''}}-7$, $\overline{x_3^{''}}=\overline{x_{3_2}}\overline{x_{3_1}}\overline{x_{3_0}}$.
\end{spacing}
\noindent $|-9x_1^{'}|_{m_2 m_3}=|-9\times 2^{2q-2}+9+9\overline{x_1^{'}}|_{9\times2^{2q-2}-1}=$ \\ $|9\overline{x_1^{'}}+8|_{9\times2^{2q-2}-1}$.

\begin{table*}[t]
\centering
\caption{The constituent bits of the $X^{'}$-expression}
\label{tab:tableI}

\resizebox{\textwidth}{!}{%
\begin{tabular}{|c|c|c|c|c|c|c|c|c|c|c|c|c|c|c|c|c|c|c|c|c|c|c|c|c|c|}
\hline
 
 & $\boldsymbol{2q}$ & $\boldsymbol{2q}$ & $\boldsymbol{2q}$ & $\boldsymbol{2q}$ & $\boldsymbol{2q}$ & $\boldsymbol{2q}$ & $\boldsymbol{2q}$ & $\boldsymbol{2q}$ & $\boldsymbol{\cdots}$ & $\boldsymbol{q}$ & $\boldsymbol{q}$ & $\boldsymbol{q}$ & $\boldsymbol{q}$ & $\boldsymbol{q}$ & $\boldsymbol{q}$ & $\boldsymbol{q}$ & $\boldsymbol{q}$ & $\boldsymbol{q}$ & $\boldsymbol{q}$ & $\boldsymbol{\cdots}$ & $\boldsymbol{4}$ & $\boldsymbol{3}$ & $\boldsymbol{2}$ & $\boldsymbol{1}$ & $\boldsymbol{0}$ \\ 
 & & $\boldsymbol{-1}$ & $\boldsymbol{-2}$ & $\boldsymbol{-3}$ & $\boldsymbol{-4}$ & $\boldsymbol{-5}$ & $\boldsymbol{-6}$ & $\boldsymbol{-7}$ & $\boldsymbol{\cdots}$ & $\boldsymbol{+3}$ & $\boldsymbol{+2}$ & $\boldsymbol{+1}$ &  & $\boldsymbol{-1}$ & $\boldsymbol{-2}$ & $\boldsymbol{-3}$ & $\boldsymbol{-4}$ & $\boldsymbol{-5}$ & $\boldsymbol{-6}$ & $\boldsymbol{\cdots}$ &  &  &  &  &  \\ 
\hline

\textbf{1} & & & & $\overline{x_{1_{2q}}}$ & $\overline{x_{1_{2q-1}}}$ & $\overline{x_{1_{2q-2}}}$ & $\overline{x_{1_{2q-3}}}$ & $\overline{x_{1_{2q-4}}}$ & & $\overline{x_{1_{q+6}}}$ & $\overline{x_{1_{q+5}}}$ & $\overline{x_{1_{q+4}}}$ & $\overline{x_{1_{q+3}}}$ & $\overline{x_{1_{q+2}}}$ & $\overline{x_{1_{q+1}}}$ & $\overline{x_{1_{q}}}$ & $\overline{x_{1_{q-1}}}$ & $\overline{x_{1_{q-2}}}$ & $\overline{x_{1_{q-3}}}$ & $\cdots$ & $\overline{x_{1_{7}}}$ & $\overline{x_{1_{6}}}$ & $\overline{x_{1_{5}}}$ & $\overline{x_{1_{4}}}$ & $\overline{x_{1_{3}}}$ \\ 
\hline
\textbf{2} & $\overline{x_{1_{2q}}}$ & $\overline{x_{1_{2q-1}}}$ & $\overline{x_{1_{2q-2}}}$ & $\overline{x_{1_{2q-3}}}$ & $\overline{x_{1_{2q-4}}}$ & $\overline{x_{1_{2q-5}}}$ & $\overline{x_{1_{2q-6}}}$ & $\overline{x_{1_{2q-7}}}$ & & $\overline{x_{1_{q+3}}}$ & $\overline{x_{1_{q+2}}}$ & $\overline{x_{1_{q+1}}}$ & $\overline{x_{1_{q}}}$ & $\overline{x_{1_{q-1}}}$ & $\overline{x_{1_{q-2}}}$ & $\overline{x_{1_{q-3}}}$ & $\overline{x_{1_{q-4}}}$ & $\overline{x_{1_{q-5}}}$ & $\overline{x_{1_{q-6}}}$ & $\cdots$ & $\overline{x_{1_{4}}}$ & $\overline{x_{1_{3}}}$ & & & \\
\hline
\textbf{3} & $x_{2_{2}}$ & $x_{2_{1}}$ & $x_{2_{0}}$ & $x_{2_{q}}$ & $1$ & $\overline{x_{2_{q}}}$ & $x_{2_{q-1}}$ & $x_{2_{q-2}}$ & & $x_{2_{8}}$ & $x_{2_{7}}$ & $x_{2_{6}}$ & $x_{2_{5}}$ & $x_{2_{4}}$ & & & $x_{2_{q}}$ & $x_{2_{q-1}}$ & $x_{2_{q-2}}$ & $\cdots$ & $x_{2_{8}}$ & $x_{2_{7}}$ & $x_{2_{6}}$ & $x_{2_{5}}$ & $x_{2_{4}}$ \\
\hline
\textbf{4} & $\overline{x_{2_{3}}}$ & $\overline{x_{2_{2}}}$ & $\overline{x_{2_{1}}}$ & & & $x_{2_{q-1}}$ & $x_{2_{q-2}}$ & $x_{2_{q-3}}$ & & $x_{2_{7}}$ & $x_{2_{6}}$ & $x_{2_{5}}$ & $x_{2_{4}}$ & $x_{2_{q}}$ & $x_{2_{q-1}}$ & $x_{2_{q-2}}$ & $x_{2_{q-3}}$ & $x_{2_{q-4}}$ & $x_{2_{q-5}}$ & $\cdots$ & $x_{2_{5}}$ & $x_{2_{4}}$ & $\overline{x_{1_{2}}}$ & $\overline{x_{1_{1}}}$ & $\overline{x_{1_{0}}}$ \\
\hline
\textbf{5} & $x_{2_{0}}$ & & $x_{2_{q}}$ & $x_{2_{q-1}}$ & $x_{2_{q-2}}$ & $x_{2_{q-3}}$ & $x_{2_{q-4}}$ & $x_{2_{q-5}}$ & & $x_{2_{5}}$ & $x_{2_{4}}$ & & & 1 & 1 & & $x_{3_{q}}$ & $x_{3_{q-1}}$ & $x_{3_{q-2}}$ & $\cdots$ & $x_{3_{8}}$ & $x_{3_{7}}$ & $x_{3_{6}}$ & $x_{3_{5}}$ & $x_{3_{4}}$ \\
\hline
\textbf{6} & & $x_{2_{q}}$ & $x_{2_{q-1}}$ & $x_{2_{q-2}}$ & $x_{2_{q-3}}$ & $x_{2_{q-4}}$ & $x_{2_{q-5}}$ & $x_{2_{q-6}}$ & & $x_{2_{4}}$ & & & & $x_{3_{q}}$ & $x_{3_{q-1}}$ & $x_{3_{q-2}}$ & $x_{3_{q-3}}$ & $x_{3_{q-4}}$ & $x_{3_{q-3}}$ & $\cdots$ & $x_{3_{5}}$ & $x_{3_{4}}$ & $x_{2_{3}}$ & $x_{2_{2}}$ & $x_{2_{1}}$ \\
\hline
\textbf{7} & & & $\overline{x_{2_{3}}}$ & & & $\overline{x_{3_{q}}}$ & $\overline{x_{3_{q-1}}}$ & $\overline{x_{3_{q-2}}}$ & & $\overline{x_{3_{8}}}$ & $\overline{x_{3_{7}}}$ & $\overline{x_{3_{6}}}$ & $\overline{x_{3_{5}}}$ & $\overline{x_{3_{4}}}$ & $x_{2_{3}}$ & $x_{2_{2}}$ & $x_{2_{1}}$ & $x_{2_{0}}$ & & & & $x_{3_{3}}$ & $x_{3_{2}}$ & $x_{3_{1}}$ & $x_{3_{0}}$ \\
 \hline
\textbf{8} & $x_{3_{3}}$ & $x_{3_{2}}$ & $x_{3_{1}}$ & & $\overline{x_{3_{q}}}$ & $\overline{x_{3_{q-1}}}$ & $\overline{x_{3_{q-2}}}$ & $\overline{x_{3_{q-3}}}$ & & $\overline{x_{3_{7}}}$ & $\overline{x_{3_{6}}}$ & $\overline{x_{3_{5}}}$ & $\overline{x_{3_{4}}}$ & $x_{2_{3}}$ & $x_{2_{2}}$ & $x_{2_{1}}$ & $x_{2_{0}}$ & & & & & & $\overline{x_{3_{3}}}$ & $\overline{x_{3_{2}}}$ & $\overline{x_{3_{1}}}$ \\
\hline
\textbf{9} & $\overline{x_{3_{0}}}$ & & $\overline{x_{3_{q}}}$ & $\overline{x_{3_{q-1}}}$ & $\overline{x_{3_{q-2}}}$ & $\overline{x_{3_{q-3}}}$ & $\overline{x_{3_{q-4}}}$ & $\overline{x_{3_{q-5}}}$ & & $\overline{x_{3_{5}}}$ & $\overline{x_{3_{4}}}$ & $x_{2_{3}}$ & $x_{2_{2}}$ & $x_{2_{1}}$ & $x_{2_{0}}$ & & 1 & & & & & & & & $x_{2_{3}}$ \\
\hline
\textbf{10} & & $\overline{x_{3_{q}}}$ & $\overline{x_{3_{q-1}}}$ & $\overline{x_{3_{q-2}}}$ & $\overline{x_{3_{q-3}}}$ & $\overline{x_{3_{q-4}}}$ & $\overline{x_{3_{q-5}}}$ & $\overline{x_{3_{q-5}}}$ & & $\overline{x_{3_{4}}}$ & $x_{2_{3}}$ & $x_{2_{2}}$ & $x_{2_{1}}$ & $x_{2_{0}}$ & $\overline{x_{3_{3}}}$ & $\overline{x_{3_{2}}}$ & $\overline{x_{3_{1}}}$ & $\overline{x_{3_{0}}}$ & 1 & $\cdots$ & 1 & 0 & 1 & 1 & 0 \\
\hline
\textbf{11} & & & & $x_{3_{3}}$ & $x_{3_{2}}$ & $x_{3_{1}}$ & $x_{3_{0}}$ & & & & & & & $\overline{x_{3_{3}}}$ & $\overline{x_{3_{2}}}$ & $\overline{x_{3_{1}}}$ & $\overline{x_{3_{0}}}$ & & & & & & & & \\
\hline
\textbf{12} & $\overline{x_{1_{2}}}$ & $\overline{x_{1_{1}}}$ & $\overline{x_{1_{0}}}$ & $\overline{x_{1_{2}}}$ & $\overline{x_{1_{1}}}$ & $\overline{x_{1_{0}}}$ & & & & & & $\overline{x_{3_{3}}}$ & $\overline{x_{3_{2}}}$ & $\overline{x_{3_{1}}}$ & $\overline{x_{3_{0}}}$ & & & & & & & & & & \\
\hline
\textbf{13} & & 1 & & $x_{2_{3}}$ & $x_{2_{2}}$ & $x_{2_{1}}$ & $x_{2_{0}}$ & & & & $\overline{x_{3_{3}}}$ & $\overline{x_{3_{2}}}$ & $\overline{x_{3_{1}}}$ & $\overline{x_{3_{0}}}$ & & & & & & & & & & & \\
\hline
\end{tabular}%
}
\end{table*}
\vspace*{-10pt}

\begin{table*}[b]
\footnotesize
\centering
\caption{Seven reduction levels represented by depth of each column and \# of utilized FAs and HAs.}
\label{tab:reduction_levels}

\resizebox{\textwidth}{!}{%
\begin{tabular}{|c|c|c|c|c|c|c|c|c|c|c|c|c|c|c|c|c|c|c|c|c|c|c|c|c|c|}
\hline
Level & Column \# & $2q$ & $2q-1$ & $2q-2$ & $2q-3$ & $2q-4$ & $2q-5$ & $2q-6$ & $2q-7$ & $\cdots$ & $q$ & $q-1$ & $q-2$ & $q-3$ & $q-4$ & $q-5$ & $q-6$ & $\cdots$ & 5 & 4 & 3 & 2 & 1 & 0 & \multirow{2}{*}{Total \# of FAs} \\
\cline{1-25} \cline{1-25}
\multirow{2}{*}{I} & Depth & 7 & 8 & 10 & 10 & 11 & 13 & 12 & 10 & $\cdots$ & 10 & 13 & 11 & 8 & 11 & 8 & 7  & $\cdots$ & 7 & 7 & 7 & 8 & 8 & 8 & \\
\cline{2-26}
 & \# FA & \cellcolor{black}\color{white}2 & \cellcolor{black}\color{white}2 & \cellcolor{black}\color{white}3 & \cellcolor{black}\color{white}3 & \cellcolor{black}\color{white}4 & \cellcolor{black}\color{white}4 & \cellcolor{black}\color{white}4 & \cellcolor{black}\color{white}3 & \cellcolor{black}\color{white}$\cdots$ & \cellcolor{black}\color{white}3 & \cellcolor{black}\color{white}4 & \cellcolor{black}\color{white}3 & \cellcolor{black}\color{white}2 & \cellcolor{black}\color{white}3 & \cellcolor{black}\color{white}2 & \cellcolor{black}\color{white}2 & \cellcolor{black}\color{white}$\cdots$ & \cellcolor{black}\color{white}2 & \cellcolor{black}\color{white}2 & \cellcolor{black}\color{white}2 & \cellcolor{black}\color{white}2 & \cellcolor{black}\color{white}2 & \cellcolor{black}\color{white}2 & $5q+8$\\
\hline

\multirow{2}{*}{II} & Depth & 5 & 7 & 7+2 & 7 & 8 & 9 & 9 & 8 & $\cdots$ & 8 & 8 & 7 & 7 & 7 & 6 & 5 & $\cdots$ & 5 & 5 & 5 & 6 & 6 & 4+2 & \\
\cline{2-26}
 & \# FA & \cellcolor{black}\color{white}1 & \cellcolor{black}\color{white}2 & \cellcolor{black}\color{white}3 & \cellcolor{black}\color{white}2 & \cellcolor{black}\color{white}2 & \cellcolor{black}\color{white}3 & \cellcolor{black}\color{white}3 & \cellcolor{black}\color{white}2 & \cellcolor{black}\color{white}$\cdots$ & \cellcolor{black}\color{white}2 & \cellcolor{black}\color{white}2 & \cellcolor{black}\color{white}2 & \cellcolor{black}\color{white}2 & \cellcolor{black}\color{white}2 & \cellcolor{black}\color{white}2 & \cellcolor{black}\color{white}1 & \cellcolor{black}\color{white}$\cdots$ & \cellcolor{black}\color{white}1 & \cellcolor{black}\color{white}1 & \cellcolor{black}\color{white}1 & \cellcolor{black}\color{white}2 & \cellcolor{black}\color{white}2 & \cellcolor{black}\color{white}2 & $3q+12$ \\
\hline

\multirow{2}{*}{III} & Depth & 5 & 6 & 5+1 & 5 & 7 & 6 & 5 & 6 & $\cdots$ & 6 & 6 & 5 & 5 & 5 & 4 & 4 & $\cdots$ & 4 & 4 & 5 & 4 & 4 & 2+1 & \\
\cline{2-26}
 & \# FA & \cellcolor{black}\color{white}1 & \cellcolor{black}\color{white}2 & \cellcolor{black}\color{white}2 & \cellcolor{black}\color{white}1 & \cellcolor{black}\color{white}2 & \cellcolor{black}\color{white}2 & \cellcolor{black}\color{white}1 & \cellcolor{black}\color{white}2 & \cellcolor{black}\color{white}$\cdots$ & \cellcolor{black}\color{white}2 & \cellcolor{black}\color{white}2 & \cellcolor{black}\color{white}1 & \cellcolor{black}\color{white}1 & \cellcolor{black}\color{white}1 & \cellcolor{black}\color{white}1 & \cellcolor{black}\color{white}1 & \cellcolor{black}\color{white}$\cdots$ & \cellcolor{black}\color{white}1 & \cellcolor{black}\color{white}1 & \cellcolor{black}\color{white}1 & \cellcolor{black}\color{white}1 & \cellcolor{black}\color{white}1 & \cellcolor{black}\color{white}1 & $3q$ \\
\hline

\multirow{2}{*}{IV} & Depth & 5 & 4 & 3+1 & 5 & 5 & 3 & 5 & 4 & $\cdots$ & 4 & 3 & 4 & 4 & 4 & 3 & 3 & $\cdots$ & 3 & 3 & 4 & 3 & 3 & 1+1 & \\
\cline{2-26}
 & \# FA & \cellcolor{black}\color{white}1 & \cellcolor{black}\color{white}1 & \cellcolor{black}\color{white}1 & \cellcolor{black}\color{white}1 & \cellcolor{black}\color{white}1 & \cellcolor{black}\color{white}1 & \cellcolor{black}\color{white}1 & \cellcolor{black}\color{white}1 & \cellcolor{black}\color{white}$\cdots$ & \cellcolor{black}\color{white}1 & \cellcolor{black}\color{white}1 & \cellcolor{black}\color{white}1 & \cellcolor{black}\color{white}1 & \cellcolor{black}\color{white}1 & \cellcolor{black}\color{white}1 & \cellcolor{black}\color{white}1 & \cellcolor{black}\color{white}$\cdots$ & \cellcolor{black}\color{white}1 & \cellcolor{black}\color{white}1 & \cellcolor{black}\color{white}1 & \cellcolor{black}\color{white}1 & \cellcolor{black}\color{white}1 & \cellcolor{black}\color{white}0 & $2q$ \\
\hline

\multirow{2}{*}{V} & Depth & 4 & 3 & 3+1 & 4 & 4 & 2 & 4 & 3 & $\cdots$ & 3 & 2 & 3 & 3 & 3 & 2 & 2 & $\cdots$ & 2 & 2 & 3 & 2 & 1 & 2+1 & \\
\cline{2-26}
 & \# FA & \cellcolor{black}\color{white}1 & \cellcolor{black}\color{white}1 & \cellcolor{black}\color{white}1 & \cellcolor{black}\color{white}1 & \cellcolor{black}\color{white}1 & \cellcolor{black}\color{white}0 & \cellcolor{black}\color{white}1 & \cellcolor{black}\color{white}1 & \cellcolor{black}\color{white}$\cdots$ & \cellcolor{black}\color{white}1 & \cellcolor{black}\color{white}0 & \cellcolor{black}\color{white}1 & \cellcolor{black}\color{white}1 & \cellcolor{black}\color{white}1 & \cellcolor{black}\color{white}0 & \cellcolor{black}\color{white}0 & \cellcolor{black}\color{white}$\cdots$ & \cellcolor{black}\color{white}0 & \cellcolor{black}\color{white}0 & \cellcolor{black}\color{white}1 & \cellcolor{black}\color{white}0 & \cellcolor{black}\color{white}0 & \cellcolor{black}\color{white}1 & $q+3$ \\
\hline

\multirow{2}{*}{VI} & Depth & 3 & 2 & 3+1 & 3 & 2 & 3 & 3 & 2 & $\cdots$ & 1 & 3 & 2 & 2 & 2 & 2 & 2 & $\cdots$ & 2 & 3 & 1 & 2 & 2 & 1+1 & \\
\cline{2-26}
 & \# FA & \cellcolor{black}\color{white}1 & \cellcolor{black}\color{white}0 & \cellcolor{black}\color{white}1 & \cellcolor{black}\color{white}1 & \cellcolor{black}\color{white}0 & \cellcolor{black}\color{white}1 & \cellcolor{black}\color{white}1 & \cellcolor{black}\color{white}0 & \cellcolor{black}\color{white}$\cdots$ & \cellcolor{black}\color{white}0 & \cellcolor{black}\color{white}1 & \cellcolor{black}\color{white}$1/2$ & \cellcolor{black}\color{white}$1/2$ & \cellcolor{black}\color{white}$1/2$ & \cellcolor{black}\color{white}$1/2$ & \cellcolor{black}\color{white}$1/2$ & \cellcolor{black}\color{white}$\cdots$ & \cellcolor{black}\color{white}$1/2$ & \cellcolor{black}\color{white}1 & \cellcolor{black}\color{white}0 & \cellcolor{black}\color{white}0 & \cellcolor{black}\color{white}0 & \cellcolor{black}\color{white}0 & $q/2+4$ \\
\hline

\multirow{2}{*}{VII} & Depth & 1 & 3 & 3+1 & 1 & 3 & 2 & 1 & 2 & $\cdots$ & 2 & 2 & 2 & 2 & 2 & 2 & 2 & $\cdots$ & 2 & 1 & 1 & 2 & 2 & 2+1 & \\
\cline{2-26}
 & \# FA & \cellcolor{black}\color{white}0 & \cellcolor{black}\color{white}1 & \cellcolor{black}\color{white}1 & \cellcolor{black}\color{white}0 & \cellcolor{black}\color{white}1 & \cellcolor{black}\color{white}0 & \cellcolor{black}\color{white}0 & \cellcolor{black}\color{white}0 & \cellcolor{black}\color{white}$\cdots$ & \cellcolor{black}\color{white}0 & \cellcolor{black}\color{white}0 & \cellcolor{black}\color{white}0 & \cellcolor{black}\color{white}0 & \cellcolor{black}\color{white}0 & \cellcolor{black}\color{white}0 & \cellcolor{black}\color{white}0 & \cellcolor{black}\color{white}$\cdots$ & \cellcolor{black}\color{white}0 & \cellcolor{black}\color{white}0 & \cellcolor{black}\color{white}0 & \cellcolor{black}\color{white}$1/2$ & \cellcolor{black}\color{white}$1/2$ & \cellcolor{black}\color{white}1 & 5 \\
\hline

\multicolumn{2}{|c|}{} & 2 & 2 & 2 & 2 & 2 & 2 & 2 & 2 & $\cdots$ & 2 & 2 & 2 & 2 & 2 & 2 & 2 & $\cdots$ & 2 & 1 & 2 & 2 & 2 & 1 &  \\
\hline

\multicolumn{25}{|c|}{\textbf{Grand total \# of FAs}} & $13.5q+32$ \\
\hline
\multicolumn{26}{|c|}{Columns $0$ and $(2q-1)$ receive seven reentrant and inverted reentrant carries, respectively. HA is considered as $1/2$ FA. } \\
\hline

\end{tabular}%
}
\end{table*}

\begin{flalign*}
X &= x_1 + 2^{2q+1}X^{'},X^{'}= &\\
\end{flalign*}
\vspace*{-25pt}
\begin{equation}
\hspace*{-8pt} 
\small 
\left|
\begin{aligned}
\label{eqn9}
\begin{split}
&(2^{q+3}+2^{q+2}+2^q+2^{q-1})(x'_2+\overline{x'_3} )+x''_3+\\
&9(\overline{x'_1}+x'_2+x'_3)+2^{2q-2} x_{2_2} x_{2_1} x_{2_0}+\\
&(2^{2q-2}+2^{2q-5}+1) \overline{x''_1}+2^{2q-6} (x''_2+x''_3)+\\
&2^{2q-2} (\overline{x_{2_3}}  \overline{x_{2_2}}  \overline{x_{2_1}}+x_{3_3} x_{3_2} x_{3_1} )+2^{2q-2} \overline{x_{2_3}}+x_{2_3}+\\
&(2^{q-1}+2^{q-2}+2^{q-4}+2^{q-5})(x''_2+\overline{x''_3})+\\
&\overline{x_{3_3}}  \overline{x_{3_2}}  \overline{x_{3_1}}+2^{2q} (x_{2_0}+\overline{x_{3_0}})+x_{2_3} x_{2_2} x_{2_1}+\\
&2^{2q-4}+2^{2q-5}+2^{q-1}+2^{q-2}+2^{q-4}+2^{q-5}-9
\end{split}
\end{aligned}
\right|_{9\times2^{2q-2}-1} 
\end{equation}


Table~\ref{tab:tableI} contains all the weighted bits of (\ref{eqn9}), organized as thirteen $(2q+1)$-bit numbers to be added modulo-$(2^{2q+1}+2^{2q-2}-1)$. 
However, the cases of $q\leq8$ require some changes before implementation (e.g., $2^{q-5}-9<0$). 
The number of required full adders (FA) are shown in Table~\ref{tab:reduction_levels} (black shaded figures, with $1/2$ for half adders), where the depth of all columns, after each of the consecutive reduction levels, are also indicated. 
However, the seven carries $c$ that spill over the leftmost column reenter the Table as in the following relation, which is appropriately reflected in Tables~\ref{tab:tableI} and \ref{tab:reduction_levels}.\\

\noindent {\small $|2^{2q+1}c|_{2^{2q+1}+2^{2q-2}-1}=(-2^{2q-2}+1)c=-2^{2q-2}+2^{2q-2} \overline{c}+c$ \\

Note that the delay overhead of the reentrant carries is only equal to that of two CSA levels, as the non-modular $13:2$ reduction requires five CSA levels. 
The final modulo-$(2^{2q+1}+2^{2q-2}-1)$ addition is handled via a modular adder, whose architecture is pretty much the same as that of modulo-$(2^q+2^{q-1}-1)$ adder of \cite{OurRNSPaper}. 
Fig.~\ref{fig:reverseConverterFigure} depicts the overall reverse converter circuitry.

\begin{figure}
    \centering
    \captionsetup{justification=centering}
    \includegraphics[width=.8\columnwidth]{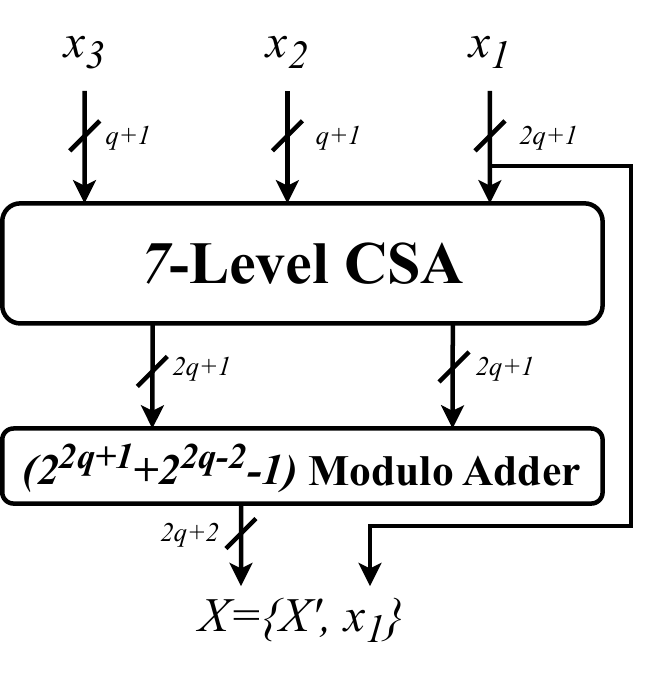}
    \caption{Reverse converter for moduli set $\tau^+$}
    \label{fig:reverseConverterFigure}
\end{figure}
\vspace*{-8pt}

\section{EVALUATION AND COMPARISON}
The comparison between two RNS moduli sets is primarily focused on the speed of modular arithmetic within the residue channels. 
This focus arises because the number of arithmetic operations performed between the initial residue generation and the final conversion back to binary is typically in the range of hundreds (such as the number of taps in an RNS-based FIR filter\cite{Application5}) or even thousands (for example, in matrix multiplications used in machine learning \cite{ref20}).
For the moduli set $\tau^+$, it has been demonstrated that the speed of modulo-$(2^q+2^{q-1}\pm 1)$ RPP adders is comparable to that of modulo-$(2^{q}\pm 1)$ adders in the $\tau={2^{q},2^{q}\pm 1}$ moduli set, as shown in previous studies \cite{OurRNSPaper, DaryushPaper}. 
However, a significant advantage of the $\tau^+$ moduli set is that its dynamic range (DR) is at least $2^q$ times greater than that of the commonly used $\tau$ moduli set for the same value of $q$ ($2^{4q+2}+2^{4q-1}-2^{2q+1}$ vs. $2^{3q}-2^{q}$).
While the forward and reverse converters for $\tau^+$ are indeed slower and costlier than those for $\tau$, this is not a significant drawback. 
The reason is that the overhead from these slower and costlier conversions is offset by the superior performance of the adders in $\tau^+$, especially as the number of operations increases. 
As a result, the overall efficiency gains in the modular arithmetic operations outweigh the additional cost and delay introduced by the conversions.
To provide a more accurate assessment, considering that the residue channels for modulo $2^q+2^{q-1}+1$ in $\tau^+$ and $2^{q^{'}}+1$ in $\tau$ are the slowest, we have calculated the delay associated with $k$ consecutive modular additions. 
These additions are preceded by a forward conversion and followed by a reverse conversion (a sequence consisting of a forward conversion, $k$ modular additions, and a final reverse conversion). 
The corresponding total delay formulas are provided in equation (\ref{eqn10}).

\[
\tau_{delay}=(18+4 \lceil \log q^{'} \rceil +k(3+2 \lceil \log q^{'} \rceil)) \Delta G,
\]
\begin{equation}
\label{eqn10}
    \tau _{delay}^+=(2q +45+4 \lceil \log q \rceil +k(4+2 \lceil \log q \rceil )) \Delta G
\end{equation}

To equalize the dynamic ranges (DR) of the corresponding moduli sets, different values of $q$ are employed for each set. 
As illustrated in Table.~\ref{tab:kComparison}, with equivalent dynamic ranges, our proposed moduli set requires fewer bits. 
As a result, after a certain threshold value of $k$, the overhead from the forward and reverse conversions in $\tau^+$ is surpassed by the performance gains achieved through the use of faster adders.
Hence, we recommend using $\tau ^+$ instead of $\tau$ for applications with $k$ greater than the turning points in Table.~\ref{tab:kComparison}, such as typical FIR filters with more than 100 taps.

\begin{table}[h]
    \centering
    \caption{Channel-widths of $\tau^+$ and $\tau$ with the same DR and the turning point of using $\tau^+$ instead of $\tau$.}
    \begin{tabular}{c|c|c|c|c} 
         \multicolumn{2}{c|}{\textbf{$\tau^+$}} & \multicolumn{2}{c|}{\textbf{$\tau$}} & Minimum $k$ for \\ \cline{1-4}
         $q$ & $\Delta G$  & $q^{'}$ & $\Delta G$ & $\tau^+$-delay $< \tau$-delay  \\ \hline
         4   & $8k+61$     & 7       & $9k+30$    & 31                            \\ \hline
         8   & $10k+73$    & 12      & $11k+34$   & 39                            \\ \hline
         16  & $12k+93$    & 23      & $13k+38$   & 55                            \\ \hline
         32  & $14k+129$   & 44      & $15k+42$   & 87                            \\  
    \end{tabular}
    \label{tab:kComparison}
\end{table}

We have also implemented the residue generators, adders, and reverse converters for the moduli-set $\tau$ and moduli-set $\tau ^+$ using Verilog. 
These designs were synthesized using the Vivado 2023.1 software on the Xilinx Artix-7 AC701 Evaluation Platform.
The findings, which also include the estimated  and power consumption for both designs, are summarized in Table.~\ref{tab:Results}.
A speedup is anticipated with higher $k$ values; for example, $8\%$ speedup is observed when $k=100$ and $q=8$ for $\tau^+$ compared to $\tau$ with an equal dynamic range and $q^{'}=12$. 
Additionally, this configuration also results in $32\%$ reduction in power dissipation.
These results are due to the smaller channel width of the adders in $\tau^+$. 
Note that the results shown in the Table.~\ref{tab:Results} correspond to specific values of $k$. 
As $k$ increases, the area deficiency remains constant; however, the improvements in delay and power become significantly more pronounced.

\begin{table}[h]
\centering
\caption{Figures of merit of forward/reverse converters $+k$ additions}
\newcolumntype{?}{!{\vrule width 3pt}}
\label{tab:Results}

\resizebox{\columnwidth}{!}{%
\begin{tabular}{|c|c|c?|c|c|c|c|c|c|}
\hline
\multirow{2}{*}{$k$} & \multirow{2}{*}{Design} & \multirow{2}{*}{$q$} & \multicolumn{2}{c|}{Delay} & \multicolumn{2}{c|}{Area}  & \multicolumn{2}{c|}{Power}  \\
\cline{4-9}
 & & & (ns) & ratio & (\# of LUTs) & ratio & ($\mu$W) & ratio\\ 
\hline
\multirow{2}{*}{39} & $\tau^+$  & 8  & 401.5  & 1.0  & 1383 & 1.0  & 246 & 1.0 \\ 
\cline{2-9}
                    & $\tau$    & 12 & 405.6  & 1.01 & 519 & 0.37  & 309 & 1.25 \\ 
\hline
\multirow{2}{*}{55} & $\tau^+$  & 16 & 774.6  & 1.0  & 4659 & 1.0  & 657 & 1.0 \\ 
\cline{2-9}
                    & $\tau$    & 23 & 783.9  & 1.01 & 1335 & 0.29 & 826 & 1.26 \\ 
\hline
\multirow{2}{*}{87} & $\tau^+$  & 32 & 1298.9 & 1.0  & 10938 & 1.0 & 1881 & 1.0 \\ 
\cline{2-9}
                    & $\tau$    & 44 & 1360.3 & 1.04 & 3772 & 0.34 & 2453 & 1.30 \\ 
\hline
\multirow{2}{*}{100} & $\tau^+$ & 8  & 911.6  & 1.0  & 1383 & 1.0  & 523 & 1.0 \\ 
\cline{2-9}
                     & $\tau$   & 12 & 987.6  & 1.08 & 519 & 0.37  & 693 & 1.32 \\ 
\hline
\end{tabular}%
}
\end{table}

\section{CONCLUSIONS AND FUTURE WORKS}
Given the design and implementation of efficient RPP modulo-$(2^q+2^{q-1}\pm1)$ adders in \cite{OurRNSPaper} and \cite{DaryushPaper}, with compatible speed to similar designs for moduli $(2^q\pm1)$, we presented the new moduli set $\tau^+=\{2^q,2^q+2^{q-1}\pm1\}$, with at least $2^q$ X DR in comparison with the popular moduli set $\tau=\{2^q,2^q\pm1\}$. 
There are several practical RNS applications that require the same channel bit-widths, but higher DR than that of $\tau$. 
Therefore, to accordingly make the new $\tau^+$ set readily available, circuit design for forward and reverse conversions and the modulo-$(2^q+2^{q-1}\pm1)$ multipliers are in order. 
Design of the required converters is presented in this work. 
Additionally, the experimental findings reveal that with an increase in $k$, enhancements in power, and delay metrics are more pronounced. 
For example, in the case where $k=100$ and $q=8$, in comparison to an equal DR $\tau$ with $q=12$, the $\tau^+$ configuration results in $8\%$ speed gain, and $32\%$ decrease in power consumption.
As for the future relevant work, we plan to conclude the ongoing research on the design and implementation of the modulo-$(2^q+2^{q-1}\pm1)$ multipliers.\\

\subsection{Acknowledgments}
Jaberipur’s research was supported by the Brain Pool program funded by the Ministry of Science and ICT through the National Research Foundation of Korea (RS-2023-00263909).\\

\appendix
\section{Auxiliary equations}
\noindent \textbf{Appendix A} (Auxiliary equations): 
\label{appendixA}

\begin{spacing}{1.3}
\begin{itemize}
  \item[\large \textcircled{\normalsize a}] $m|Z|_{m'}=|mZ|_{mm'}$ since $m|Z|_{m'}= \\ m
        \begin{cases}
             Z     & \text{if } Z<m' \\
             Z-m'  & \text{if } Z\geq m'
       \end{cases} = \\
       \begin{cases}
             mZ     & \text{if } mZ<mm' \\
             mZ-mm' & \text{if } mZ\geq mm'
       \end{cases}$
       
  \item[\large \textcircled{\normalsize b}] $|\mu _2 m_3 |_{m_2}=1 \Rightarrow \mu _2=3\times 2^{q-2}$ 
  \item[\large \textcircled{\normalsize c}] $2\mu _2=3\times 2^{q-1}=m_2+1$
  \item[\large \textcircled{\normalsize d}] $x_2=|X|_{m_2}=16{x'_2}+x''_2, x'_2=x_{2_q}\cdots x_{2_4}, x''_2=x_{2_3} x_{2_2} x_{2_1} x_{2_0}$
  \item[\large \textcircled{\normalsize e}] $x_3=|X|_{m_3}=16x'_3+x''_3, x'_3=x_{3_q}\cdots x_{3_4}, x''_3=x_{3_3} x_{3_2} x_{3_1} x_{3_0}$
  \item[\large \textcircled{\normalsize f}] $|\mu _1 m_1 |_{m_2 m_3}=1 \Rightarrow$ \\ $\mu _1=2^{2q-2}+2^{2q-5}+1=9\times 2^{2q-5}+1$
  \item[\large \textcircled{\normalsize g}] $|8\mu _1|_{m_2 m_3}=|9\times 2^{2q-2}+8|_{9\times 2^{2q-2}-1}=9$ 
  \item[\large \textcircled{\normalsize h}] $\mu _1=3\times 2^{q-4}\times 3\times 2^{q-1}+1=3\times 2^{q-4} (m_2+1)+1$
\end{itemize}
\end{spacing}

\section{Derivation of (\ref{eqn9})}
\noindent \textbf{Appendix B} (Derivation of (\ref{eqn9})): \\
\label{appendixB}

\noindent$X'=|X''+X''_{23}|_{m_2 m_3}=$ \\
\vspace*{-10pt}
\begin{flalign*}
\left|
\begin{aligned}
\begin{split}
&9x'_3 + \mu _1 x''_3 - 9x'_1 - \mu _1 x''_1 + 9m_3 (x'_2 - x'_3)+ \\
&(3\times 2^{q-4} + 1) m_3 (x''_2 - x''_3)
\end{split}
\end{aligned}
\right|_{m_2 m_3} = &&
\end{flalign*}

\vspace*{-10pt}

\begin{flalign*}
\left|
\begin{aligned}
\begin{split}
&-x'_3 \times 9(m_3-1) + 9\overline{x'_1}+ 9 \times 2^{2q-5}(-7+\overline{x''_1})+\\
&(9 \times 2^{2q-5} + 1 - (3 \times 2^{2q-4} + 1)(3 \times 2^{q-1}+1))x''_3\\
&+ (3 \times 2^{q-4}+1)(3 \times 2^{q-1}+1)x''_2\\
&+8-7 +\overline{x''_1} +9m_3x'_2 
\end{split}
\end{aligned}
\right|_{m_2m_3}=&&
\end{flalign*}

\vspace*{-10pt}

\begin{flalign*}
\left|
\begin{aligned}
\begin{split}
&27 \times 2^{q-1} (\overline{x'_3}+1)+9 \times 2^{2q-3}-2-\\
&27 \times 2^{q-4} x''_3+9\overline{x'_1}+8-63 \times 2^{2q-5}+\\
&9 \times 2^{2q-5} \overline{x''_1}-7+\overline{x''_1}+27 \times 2^{q-1} x'_2+9x'_2+\\
&(9 \times 2^{2q-5}+3 \times 2^{q-4}+3 \times 2^{q-1}+1) x''_2
\end{split}
\end{aligned}
\right|_{m_2 m_3}=&&
\end{flalign*}

\vspace*{-10pt}

\begin{flalign*}
\left|
\begin{aligned}
\begin{split}
&27 \times 2^{q-1} \overline{x'_3}+27 \times 2^{q-4} \overline{x''_3}-7 \times 27 \times 2^{q-4}+\\
&9\overline{x'_1}+9 \times 2^{2q-5} \overline{x''_1}+\overline{x''_1}+27 \times 2^{q-1} x'_2+9x'_2+\\
&(9 \times 2^{2q-5}+27 \times 2^{q-4})x''_2 +x''_2+27 \times 2^{q-1}+\\
&9 \times 2^{2q-3}-2+8-63 \times 2^{2q-5}-7
\end{split}
\end{aligned}
\right|_{m_2 m_3}=&&
\end{flalign*}

\vspace*{-10pt}

\begin{flalign*}
\left|
\begin{aligned}
\begin{split}
&9(\overline{x'_1}+x'_2)+27 \times 2^{q-1} (\overline{x'_3}+x'_2)+\overline{x''_1}+\\
&x''_2 9 \times 2^{2q-5} (\overline{x''_1}+x''_2)+27 \times 2^{q-4} (x''_2+\overline{x''_3})+\\
&27 \times 2^{q-4}-27 \times 2^{2q-5}-1+9 \times 2^{2q-2}-1
\end{split}
\end{aligned}
\right|_{m_2 m_3}=&&
\end{flalign*}

\vspace*{-10pt}

\begin{flalign*}
\left|
\begin{aligned}
\begin{split}
&8\overline{x'_1}+\overline{x''_1}+8x'_2+x''_2+9 \times 2^{2q-5} (\overline{x''_1}+x''_2)+\overline{x'_1}+\\
&x'_2+27 \times 2^{q-4} (8\overline{x'_3}+\overline{x''_3}+8x'_2+x''_2)+27 \times 2^{q-4}\\
&+45 \times 2^{2q-5}-2
\end{split}
\end{aligned}
\right|_{m_2 m_3}=&&
\end{flalign*}

\vspace*{-10pt}

\begin{flalign*}
\left|
\begin{aligned}
\begin{split}
&\overline{x'_1}+x'_2+\overline{x_1}+x_2+27 \times 2^{q-4} (x_2+\overline{x_3})+\\
&9 \times 2^{2q-5} (\overline{x''_1}+x''_2)+2^{2q}+2^{2q-2}+2^{2q-3}+2^{2q-5}\\
&+2^q+2^{q-1}+2^{q-3}+2^{q-4}-2
\end{split}
\end{aligned}
\right|_{m_2 m_3}\Rightarrow&&
\end{flalign*}

\vspace*{-14pt}
\begin{flalign*}
&X=x_1+&&\\
&2^{2q+1}
\left|
\begin{aligned}
\begin{split}
&\overline{x'_1}+x'_2+\overline{x_1}+x_2+2^{2q}+2^{2q-2}+\\
&(2^q+2^{q-1}+2^{q-3}+2^{q-4})(x_2+\overline{x_3})+\\
&(2^{2q-2}+2^{2q-5})(\overline{x''_1}+x''_2)+2^{2q-3}+2^{2q-5}+\\
&2^q+2^{q-1}+2^{q-3}+2^{q-4}-2
\end{split}
\end{aligned}
\right|_{9 \times 2^{2q-2}-1}&&
\end{flalign*}


\end{document}